\documentclass[10pt,conference]{IEEEtran}
\IEEEoverridecommandlockouts
\usepackage{cite}
\usepackage{amsmath,amssymb,amsfonts}
\usepackage{algorithmic}
\usepackage{graphicx}
\usepackage{textcomp}
\usepackage{xcolor}
\def\BibTeX{{\rm B\kern-.05em{\sc i\kern-.025em b}\kern-.08em
    T\kern-.1667em\lower.7ex\hbox{E}\kern-.125emX}}
    
\usepackage{tikz}
\usepackage{enumitem}
\usepackage{booktabs}
\usepackage{multirow}

\usepackage{xcolor}
\usepackage{colortbl}

\usepackage{pifont}

\usepackage{tikz}

\newcommand*\blackcircled[1]{\tikz[baseline=(char.base)]{
            \node[shape=circle,draw,inner sep=0.8pt,fill=black,text=white] (char) {#1};}}

\usepackage{cite}
\usepackage{amsmath,amssymb,amsfonts}
\usepackage{algorithmic}
\usepackage{graphicx}
\usepackage{textcomp}
\usepackage{colortbl}  
\usepackage{xcolor}
\usepackage{multirow} 
\usepackage{booktabs} 
\usepackage{hyperref}
\usepackage{algorithm}
\usepackage{tcolorbox}
\usepackage{tikz}
\usepackage{threeparttable}

\usepackage{enumitem}
\usepackage{fmtcount}

\usepackage{latexsym}
\usepackage{url}
\usepackage{arydshln}

\def\BibTeX{{\rm B\kern-.05em{\sc i\kern-.025em b}\kern-.08em
    T\kern-.1667em\lower.7ex\hbox{E}\kern-.125emX}}

\usepackage{xspace}

\newcommand{\mytextsuperscript}[1]{\textsuperscript{#1}}

\newcommand{\aligncoder}{{AlignCoder}\xspace}
\newcommand{\alignretriever}{AlignRetriever\xspace}

\newcommand{\ReACCCLSeven}{ReACC\textsubscript{ CodeLlama-7B}}
\newcommand{\RepocoderCLSeven}{RepoCoder\textsubscript{ CodeLlama-7B}}
\newcommand{\RlcoderCLSeven}{RLCoder\textsubscript{ CodeLlama-7B}}
\newcommand{\AligncoderCLSeven}{AlignCoder\textsubscript{ CodeLlama-7B}}

\newcommand{\ReACCSCSeven}{ReACC\textsubscript{ StarCoder-7B}}
\newcommand{\RepocoderSCSeven}{RepoCoder\textsubscript{ StarCoder-7B}}
\newcommand{\RlcoderSCSeven}{RLCoder\textsubscript{ StarCoder-7B}}
\newcommand{\AligncoderSCSeven}{AlignCoder\textsubscript{ StarCoder-7B}}

\newcommand{\ReACCSCTSeven}{ReACC\textsubscript{ StarCoder2-7B}}
\newcommand{\RepocoderSCTSeven}{RepoCoder\textsubscript{ StarCoder2-7B}}
\newcommand{\RlcoderSCTSeven}{RLCoder\textsubscript{ StarCoder2-7B}}
\newcommand{\AligncoderSCTSeven}{AlignCoder\textsubscript{ StarCoder2-7B}}

\newcommand{\ReACCDSCOne}{ReACC\textsubscript{ DeepSeekCoder-1B}}
\newcommand{\RepocoderDSCOne}{RepoCoder\textsubscript{ DeepSeekCoder-1B}}
\newcommand{\RlcoderDSCOne}{RLCoder\textsubscript{ DeepSeekCoder-1B}}
\newcommand{\AligncoderDSCOne}{AlignCoder\textsubscript{ DeepSeekCoder-1B}}

\newcommand{\ReACCDSCSeven}{ReACC\textsubscript{ DeepSeekCoder-7B}}
\newcommand{\RepocoderDSCSeven}{RepoCoder\textsubscript{ DeepSeekCoder-7B}}
\newcommand{\RlcoderDSCSeven}{RLCoder\textsubscript{ DeepSeekCoder-7B}}
\newcommand{\AligncoderSeven}{AlignCoder\textsubscript{ DeepSeekCoder-7B}}

\newcommand{\sam}[1]{Sampling Number=#1}
\newcommand{\revised}[1]{{\color{black} #1}}

\newcommand{\boxmargin}{1mm}
\tcbuselibrary{most, skins, breakable, theorems}
\newtcolorbox{myboxa}[2][]{
    colback=gray!10!white,
    colframe=black, enhanced,
    attach boxed title to top left={yshift=-2mm,xshift=5mm},
    title=#2,#1
}
\newtcolorbox{myboxb}[2][]{
    boxsep=3pt,
    left = \boxmargin, right = \boxmargin, top = \boxmargin, bottom = \boxmargin,
    title={#2},#1
}
\newtcolorbox{myboxc}{
    colback=gray!15!white,
    arc = 0pt, outer arc = 0pt,
    boxsep=0pt, left = 3pt, right = 0pt, top = 0pt, bottom = 0pt, 
    leftrule=3pt, bottomrule=0pt,toprule=0pt, rightrule=0pt,
    left = \boxmargin, right = \boxmargin, top = \boxmargin, bottom = \boxmargin
}
\newtcolorbox{myboxd}{
    colback=gray!10,
    colframe=black,
    width=\columnwidth,
    arc=1mm, auto outer arc,
    boxrule=0.5pt,
}

\begin{document}

\title{AlignCoder: Aligning Retrieval with Target Intent for Repository-Level Code Completion}

\author{
    \IEEEauthorblockN{Tianyue Jiang$^{1\dagger}$, Yanli Wang$^{1\dagger}$, Yanlin Wang$^{1*}$\thanks{* Yanlin Wang is the corresponding author, wangylin36@mail.sysu.edu.cn.}\thanks{$\dagger$ These authors contributed equally to this work.}, Daya Guo$^{3}$, Ensheng Shi$^{2}$, Yuchi Ma$^{2}$, Jiachi Chen$^{1}$, Zibin Zheng$^{1}$}
    \IEEEauthorblockA{$^{1}$ Sun Yat-sen University, Zhuhai, China}
    \IEEEauthorblockA{$^{2}$ Huawei Cloud Computing Technologies Co., Ltd., Shenzhen, China}
    \IEEEauthorblockA{$^{3}$ Independent Researcher, China}
}

\maketitle

\begin{abstract}
Repository-level code completion remains a challenging task for existing code large language models (code LLMs) due to their limited understanding of repository-specific context and domain knowledge. While retrieval-augmented generation (RAG) approaches have shown promise by retrieving relevant code snippets as cross-file context, they suffer from two fundamental problems: misalignment between the query and the target code in the retrieval process, and the inability of existing retrieval methods to effectively utilize the inference information. To address these challenges, we propose \aligncoder, a repository-level code completion framework that introduces a query enhancement mechanism and a reinforcement learning based retriever training method. Our approach generates multiple candidate completions to construct an enhanced query that bridges the semantic gap between the initial query and the target code. Additionally, we employ reinforcement learning to train an AlignRetriever that learns to leverage inference information in the enhanced query for more accurate retrieval. We evaluate \aligncoder on two widely-used benchmarks (CrossCodeEval and RepoEval) across five backbone code LLMs, demonstrating an 18.1\% improvement in EM score compared to baselines on the CrossCodeEval benchmark. The results show that our framework achieves superior performance and exhibits high generalizability across various code LLMs and programming languages.

\end{abstract}


\section{Introduction}
Recent developments in code large language models (code LLMs)~\cite{lozhkov2024starcoder, roziere2023code, guo2024deepseek, hui2024qwen2} have demonstrated impressive capability in general code completion tasks~\cite{zan2022large, zhang2023unifying, izadi2024language, izadi2024language, guo2023longcoder, ren2024codeattack, shrivastava2023repofusion}. However, existing code LLMs demonstrate suboptimal performance on repository-level code completion tasks, primarily due to their insufficient understanding of repository-specific context and domain knowledge~\cite{tang2023domain}.
This limitation stems from the fact that target code repositories are often newly created, proprietary, or work-in-progress projects, making it 
hard
for code LLMs to acquire repository-specific knowledge during pre-training and fine-tuning phases~\cite{liu2024graphcoder}.
To address this challenge, one straightforward approach leverages the increasing context window length of modern models by concatenating all repository files into a single prompt. However, this naive concatenation introduces substantial irrelevant information that interferes with model generation~\cite{shi2023large, yoran2023making}. Consequently, recent methods have adopted the retrieval-augmented generation (RAG) paradigm~\cite{lu2022reacc, zhang2023repocoder, liu2024graphcoder, phan2024repohyper, wang2024repogenreflex, bui2024rambo, liang2024repogenix}, which uses unfinished code in the current file as a query to retrieve relevant code snippets from the entire repository. These retrieved code snippets serve as cross-file context and are concatenated with the unfinished code to construct prompts for code LLMs.
For instance, ReACC~\cite{lu2022reacc} integrates both sparse and dense retrieval methods. Sparse retrievers, such as BM25~\cite{robertson2009probabilistic}, employ keyword matching algorithms that effectively capture lexical information. Conversely, dense retrievers encode both queries and code snippets into dense vectors, enabling the identification of semantically similar code snippets through vector similarity measurements.
Despite these advances, most dense retrieval methods fail to leverage the reasoning and understanding capabilities of code LLMs to enhance the retrieval process, resulting in a semantic gap between query and target code in the retrieval process. To mitigate this misalignment, RepoCoder~\cite{zhang2023repocoder} proposes an iterative retrieval strategy where the generator produces intermediate completions based on retrieved code snippets, which are then incorporated into subsequent retrieval queries. While this approach represents a significant step toward addressing the query-target misalignment, the fundamental issue remains unresolved. Specifically, we have identified the following problems in existing retrieval processes:

\begin{enumerate}[label={\bfseries P\arabic*}]
\item \textbf{The misalignment between query and target code remains unresolved.} 

To address the problem of misalignment between query and target code in RAG-based code completion, RepoCoder~\cite{zhang2023repocoder} introduces an iterative retrieval approach that concatenates the completion with the unfinished code to obtain a new query for the next retrieval round.
However, this method has two problems: If LLMs produce an incorrect completion in an iteration, it will cause chain errors that affect the subsequent retrieval. Besides, multiple retrieval rounds significantly reduce efficiency.
\item \textbf{Existing retrieval methods lack the ability to learn how to utilize inference information.}

Although RepoCoder has demonstrated that the generated completion can significantly assist repository-level code completion, the method employs sparse (e.g., Jaccard index~\cite{jaccard1912distribution}) or dense (e.g., UniXcoder~\cite{guo2022unixcoder}) retrievers that are not specifically trained. These retrievers may fail to understand the relationship between the unfinished code and the candidate completion, and therefore cannot fully leverage it for effective retrieval.
\end{enumerate}

\begin{figure*}[t]
\centering
\includegraphics[width=\linewidth]{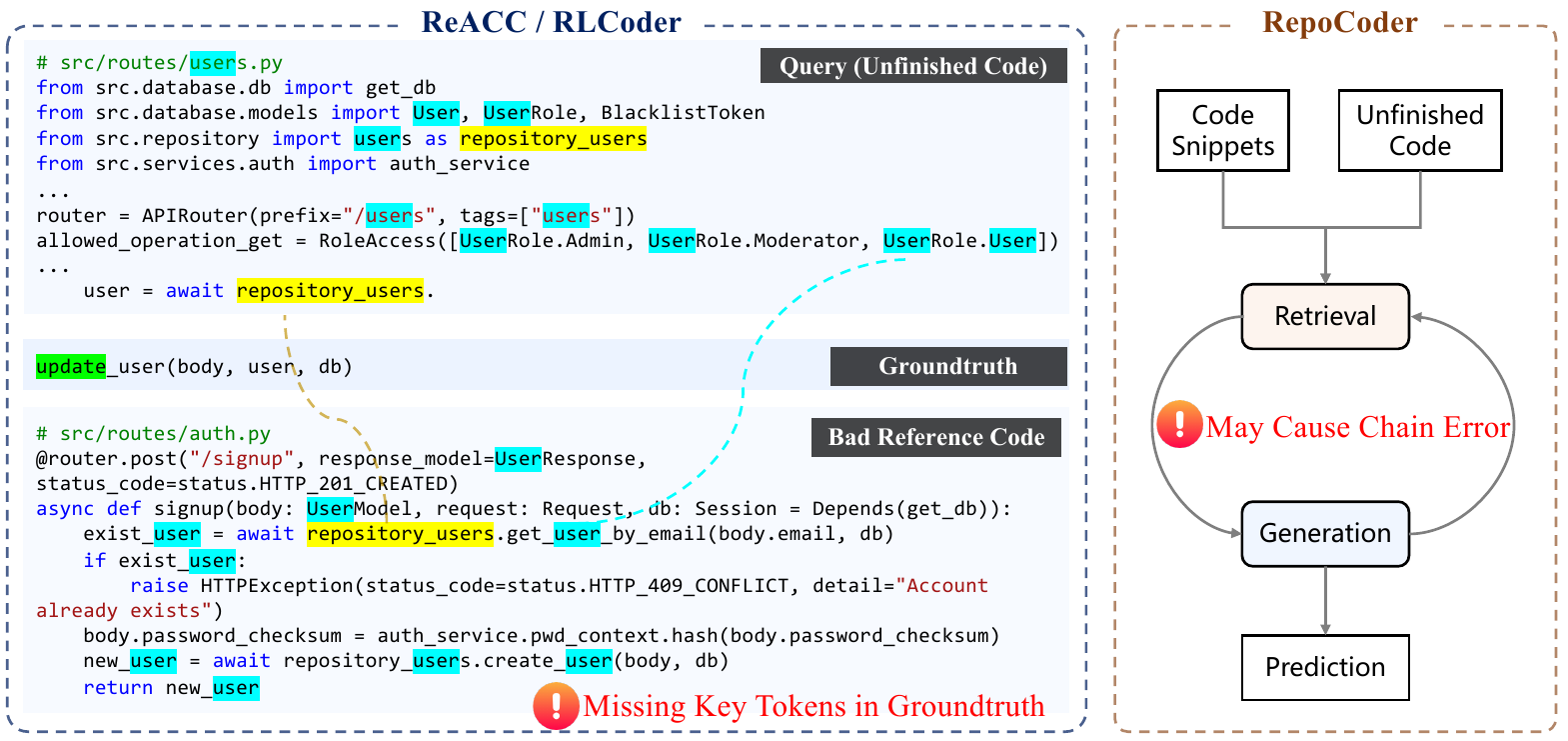}
\vspace{-20pt}
\caption{Motivating example. Limitations of prior works: ReACC and RLCoder use the unfinished code as the query, which may miss key tokens in the ground truth. RepoCoder may cause chain errors and has potential efficiency issue. }
\label{fig:MotivatingExample}
\end{figure*}

In this paper, we propose a repository-level code completion framework \aligncoder to address the two aforementioned problems. First, we introduce a query enhancement mechanism that leverages sampled candidate completions to improve retrieval accuracy. 
Specifically, our approach employs the sampler to generate multiple candidate completions, which are then expanded to the unfinished code to construct an enhanced query representation. This enhanced query effectively bridges the gap between the initial query and the desired target completion (addressing \textbf{P1}). 
Secondly, we employ reinforcement learning to train the retriever, enabling it to learn how to utilize the multiple candidate completions contained in the enhanced query for more accurate retrieval (addressing \textbf{P2}). Specifically, given an enhanced query, we employ the retriever to retrieve multiple potentially relevant code snippets. We then utilize a reward model to evaluate the perplexity (PPL) of generating the target code using these retrieved code snippets, deriving rewards from this evaluation process to update the retriever's parameters.

We evaluate \aligncoder with extensive experiments using five backbone LLMs on two benchmarks: CrossCodeEval~\cite{ding2023crosscodeeval} and RepoEval~\cite{zhang2023repocoder}. These two benchmarks are widely used in repository-level code completion. 
Experimental results show that our framework achieves an 18.1\% improvement in EM score compared with baselines on the CrossCodeEval Python. \aligncoder demonstrates high generalizability, showing effectiveness across various code LLMs and programming languages.

To summarize, our main contributions are:
\begin{itemize}
    \item We introduce \aligncoder, a repository-level code completion framework. The proposed query enhancement mechanism allows the enhanced query to have a greater possibility of including key tokens relevant to the target code. This framework effectively addresses the misalignment problem between query and target code in the retrieval process.
    \item We train the retriever using reinforcement learning, resulting in Alignretriever. This retriever learns to leverage the inference information in enhanced queries to achieve more accurate retrieval.
    \item We perform extensive experimental evaluation on various benchmarks and code LLMs. The results show that \aligncoder achieves superior performance compared to previous approaches. We provide our code and data at https://github.com/DeepSoftwareAnalytics/AlignCoder.
\end{itemize}

\section{Preliminaries}
\subsection{Retrieval-Augmented Code Completion Paradigms}

Retrieval-augmented methods have been widely adopted for repository-level code completion. Such methods can be categorized into two primary paradigms:

\textit{1) Retrieve-then-Generate:} This approach first retrieves relevant code snippets from the codebase using the unfinished code as a query, then uses the retrieved code snippets to assist the code completion. Methods like ReACC~\cite{lu2022reacc} and RLCoder~\cite{wang2024rlcoder} follow this paradigm. The retrieval process takes the unfinished code $C_u$ as input to search through the repository $R$ for the top-$k$ most similar code snippets $S$, which are then provided to the language model $P_\theta$ as context for generating the target completion $\hat{C}_t$.
However, this paradigm faces the challenge of semantic misalignment. These methods only use unfinished code as a query, which may not contain the key tokens related to the target code. Consequently, it becomes difficult to retrieve the relevant code snippets needed to generate the target code during the retrieval process~\cite{zhang2023repocoder}.

\textit{2) Iterative Generate-and-Retrieve:} This approach alternates between generation and retrieval in multiple iterations. Starting with an initial retrieval using the unfinished code, the method generates a candidate completion and then uses the concatenation of the unfinished code and the generated candidate as a new query for the next retrieval iteration. Methods like RepoCoder~\cite{zhang2023repocoder} employ this iterative approach, where each iteration refines both the retrieved context and the generated completion.
However, since each iteration relies on a single candidate completion, errors can propagate through the chain. If errors occur in the intermediate generation step or key tokens are still missing, the subsequent retrieval will be based on flawed information, leading to cascading errors throughout the remaining iterations.

\subsection{Semantic Gap in Repository Code Retrieval}

Repository-level code completion leverages contextual information across multiple files to generate accurate code completions. Formally, given an unfinished code $C_u$ and a repository $R$, the objective is to generate target code $C_t$ such that $C = C_u \oplus C_t$ is syntactically and semantically correct. A challenge in this task stems from the semantic gap between unfinished code and target code. Using $C_u$ directly as a query is suboptimal because unfinished code and target code belong to different semantic spaces, making alignment difficult~\cite{wang2023query2docqueryexpansionlarge}. The semantic gap can be formalized as :
\begin{equation}
\mathcal{G}(C_u, C_r) = d(\Phi_{query}(C_u), \Phi_{target}(C_r))
\end{equation}
where $\Phi_{query}$ and $\Phi_{target}$ be embedding functions mapping code to semantic spaces and $d$ is a distance function in the semantic space. Previous approaches like RLCoder~\cite{wang2024rlcoder} attempt to bridge this gap through reinforcement learning, while RepoCoder~\cite{zhang2023repocoder} uses iterative retrieval-generation to update the query. However, the former requires semantic space alignment between unfinished code and target code, which is inherently challenging due to the differences in their representations. The latter relies on a single sample from each iteration, which can easily propagate errors in a chain reaction—if one iteration produces an incorrect completion, subsequent iterations are built upon this flawed foundation, leading to compounding errors throughout the retrieval-generation process.

\subsection{Multiple Sampling in LLMs}

LLMs exhibit inherent stochasticity in their generation process, making single-sample outputs unreliable. The probabilistic nature of token sampling introduces variability that can lead to inconsistent or incorrect responses, even when queried with identical prompts. However, through the multiple sampling strategy, we can significantly improve the likelihood of obtaining correct completions.

\subsubsection{Single Sample Unreliability}

For a given query $q$, an LLM generates a response by sampling from the learned probability distribution over the vocabulary $V$ at each timestep. Let $p_\theta(y|x)$ denote the probability of generating sequence $y$ given input $x$ and model parameters $\theta$. The sampling process can be formulated as:
\begin{equation}
y_t \sim p_\theta(\cdot | x, y_{<t})
\end{equation}
where $y_t$ represents the token at position $t$, and $y_{<t}$ denotes all previously generated tokens.

Due to the stochastic nature of sampling, a single generation attempt has probability $p_s$ of producing a correct answer, typically $p_s < 1$ and may be lower for complex queries. Threrefore, the probability of generating an incorrect response in a single attempt is:
\begin{equation}
P(\text{error}) = 1 - p_s
\end{equation}

\subsubsection{Multiple Sampling Benefits}

When performing multiple sampling attempts, we acknowledge that individual samples are not truly independent events, as they originate from the same model with identical parameters and input prompt. However, due to the stochastic nature of the sampling process (temperature-based sampling, top-k, or nucleus sampling), we can approximate the samples as quasi-independent for analytical purposes.

Let $\rho$ denote the correlation coefficient between samples, where $0 \leq \rho \leq 1$. For truly independent samples ($\rho = 0$), the probability of obtaining at least one correct answer from $n$ attempts would be:
\begin{equation}
P_{\text{independent}}(\text{at least one correct}) = 1 - (1 - p_s)^n
\end{equation}

However, accounting for inter-sample correlation, the actual probability can be approximated as:
\begin{equation}
P(\text{at least one correct}) \approx 1 - (1 - p_s)^{n \cdot (1-\rho)}
\end{equation}

where the effective number of independent samples is reduced by the correlation factor $(1-\rho)$.

In practice, modern sampling techniques (such as temperature scaling $T > 0$ or top-p sampling) introduce sufficient randomness that $\rho$ remains relatively small, making the independence approximation reasonable:
\begin{equation}
P(\text{at least one correct}) \approx 1 - (1 - p_s)^n \quad \text{when } \rho \ll 1
\end{equation}

This relationship demonstrates that even with correlated samples, the success probability increases substantially with the number of sampling attempts. For practical applications, multiple sampling strategy consistently yields higher accuracy than single sampling, establishing it as an effective strategy for improving LLM reliability in critical applications.

\begin{figure}[t]
\centering
\includegraphics[width=0.8\columnwidth]{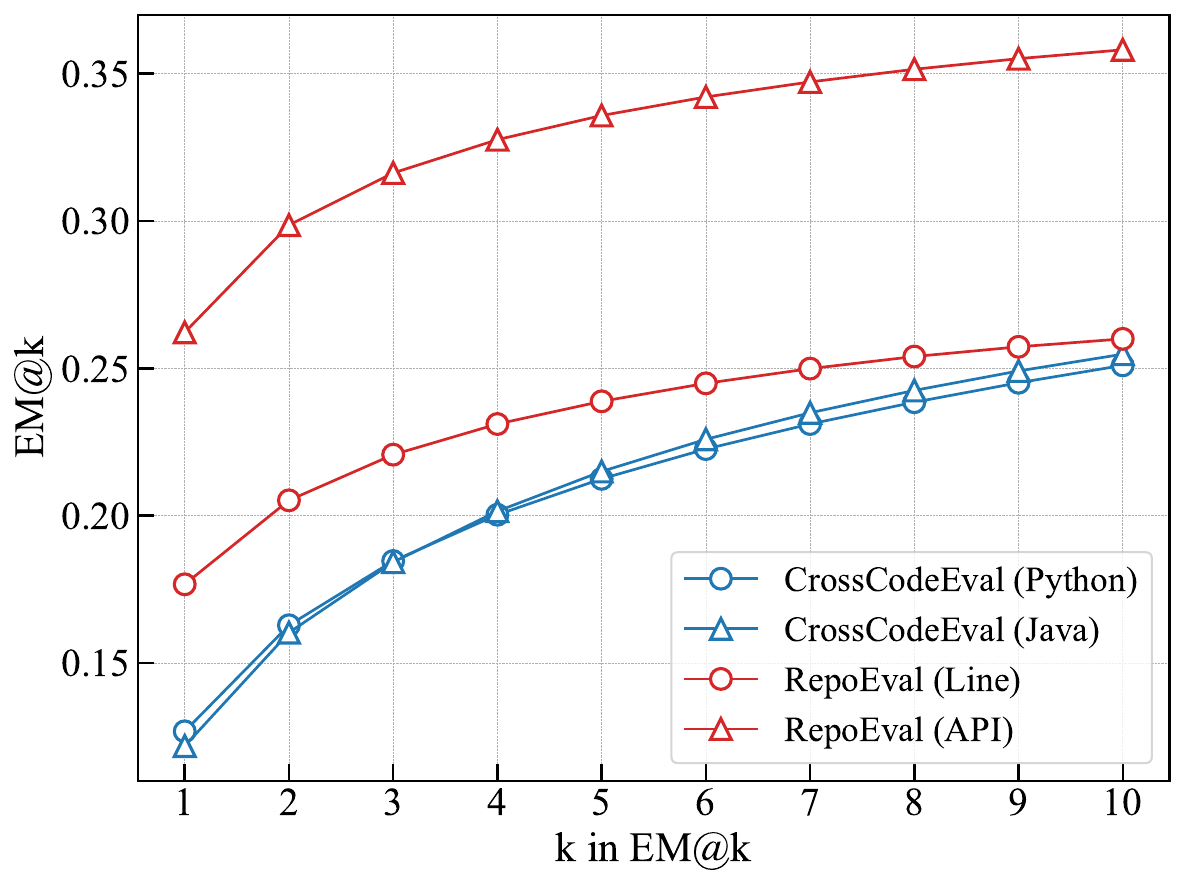}
\caption{EM@k performance trends on CrossCodeEval and RepoEval with varying k parameters. The trends indicate that multiple sampling strategy provides a higher probability of producing a correct completion compared tosingle sampling.} 
\label{fig:em_at_k}
\end{figure}

The above provides an introduction to the theoretical foundations. In the following part, we demonstrate the effectiveness of multiple sampling strategy compared to single sampling through a preliminary experiment. This experiment utilizes the CrossCodeEval and RepoEval, where the prompt consists of BM25 retrieval code snippets concatenated with unfinished code. We present the \textit{EM@k} results (k ranging from 1 to 10) for DeepSeekCoder-1B using the aforementioned prompt with 10 sampling attempts. Specifically, \textit{EM@k} represents the probability of that at least one of the k samples exactly matches the target code (EM=1). As shown in figure~\ref{fig:em_at_k}, the \textit{EM@k} values on both benchmarks increase with the value of k. This indicates that multiple sampling strategy provides a higher probability of generating a correct completion compared to single sampling, which also means it is easier to contain key tokens related to the correct completion.

\subsubsection{Diminishing Returns and Optimal Sampling Threshold}
While multiple sampling generally improves the probability of obtaining correct completions, excessive sampling can introduce diminishing returns and potentially counterproductive effects. 
As the number of samples increases beyond an optimal threshold, several factors contribute to degraded performance:

\textbf{Error Accumulation:} With increased sampling numbers, the absolute number of incorrect responses grows, potentially overwhelming correct completions during aggregation or selection processes. If we define $\epsilon_n$ as the cumulative error rate after $n$ samples, we have:
\begin{equation}
\epsilon_n = n \cdot (1 - p_s)
\end{equation}

\textbf{Selection Complexity:} The probability of selecting the correct answer from a pool containing both correct and incorrect responses depends on the selection mechanism. For random selection from $n$ samples, where $k$ samples are correct, the probability of selecting a correct answer is:
\begin{equation}
P(\text{correct selection}) = \frac{k}{n} = \frac{n \cdot p_s}{n} = p_s
\end{equation}

However, for more sophisticated selection mechanisms (e.g., majority voting, confidence-based selection), the relationship becomes more complex.

\textbf{Optimal Sampling Threshold:} To determine the optimal number of samples $n^*$, we must balance the benefit of increased correct sample probability against the cost of error accumulation. The expected utility can be formulated as:
\begin{equation}
U(n) = \alpha \cdot P(\text{at least one correct}) - \beta \cdot \epsilon_n - \gamma \cdot n
\end{equation}
where $\alpha$, $\beta$, and $\gamma$ represent the weights for correctness benefit, error penalty, and computational cost, respectively.

Substituting our previous formulations:
\begin{equation}
U(n) = \alpha \cdot [1 - (1 - p_s)^n] - \beta \cdot n(1 - p_s) - \gamma \cdot n
\end{equation}

Taking the derivative with respect to $n$ and setting it to zero:
\begin{equation}
\frac{dU(n)}{dn} = \alpha \cdot (1 - p_s)^n \ln(1 - p_s) - \beta(1 - p_s) - \gamma = 0
\end{equation}

The optimal sampling threshold $n^*$ can be approximated by solving:
\begin{equation}
n^* \approx \frac{\ln\left(\frac{\beta(1-p_s) + \gamma}{\alpha \ln(1-p_s)}\right)}{\ln(1-p_s)}
\end{equation}


This theoretical framework demonstrates that while multiple sampling is beneficial, there exists an optimal threshold beyond which additional samples provide marginal utility and may even degrade overall system performance due to increased complexity in answer selection and computational overhead.

\subsection{Perplexity in LLMs for Code Assessment}

Perplexity (PPL) quantifies how well a probability model predicts token sequences, serving as an intrinsic measure of model confidence in Large Language Models. For a token sequence $\mathbf{y} = (y_1, y_2, \ldots, y_T)$, perplexity is defined as:
\begin{equation}
\text{PPL}(\mathbf{y}) = \exp\left(-\frac{1}{T} \sum_{t=1}^{T} \log p_\theta(y_t | y_{<t})\right)
\end{equation}

where $p_\theta(y_t | y_{<t})$ represents the conditional probability of token $y_t$ given preceding context $y_{<t}$. Lower perplexity indicates higher model confidence.

While not directly measuring correctness, perplexity correlates with code quality through several mechanisms: (1) \textit{Statistical regularity:} well-formed code follows learned patterns from training data; (2) \textit{Syntactic consistency:} valid syntax exhibits predictable token transitions; (3) \textit{Semantic coherence:} logical code patterns yield lower perplexity.

However, perplexity has limitations: uncommon but correct coding styles, domain-specific patterns, and model understanding gaps can cause discrepancies between perplexity and actual correctness. Empirical studies show moderate positive correlation between low perplexity and code correctness~\cite{chen2021evaluating, austin2021program}, making it effective as a first-order heuristic for ranking code completion candidates when combined with other validation methods.


\section{Methodology}

\begin{figure*}[t]
\centering
\includegraphics[width=0.9\linewidth]{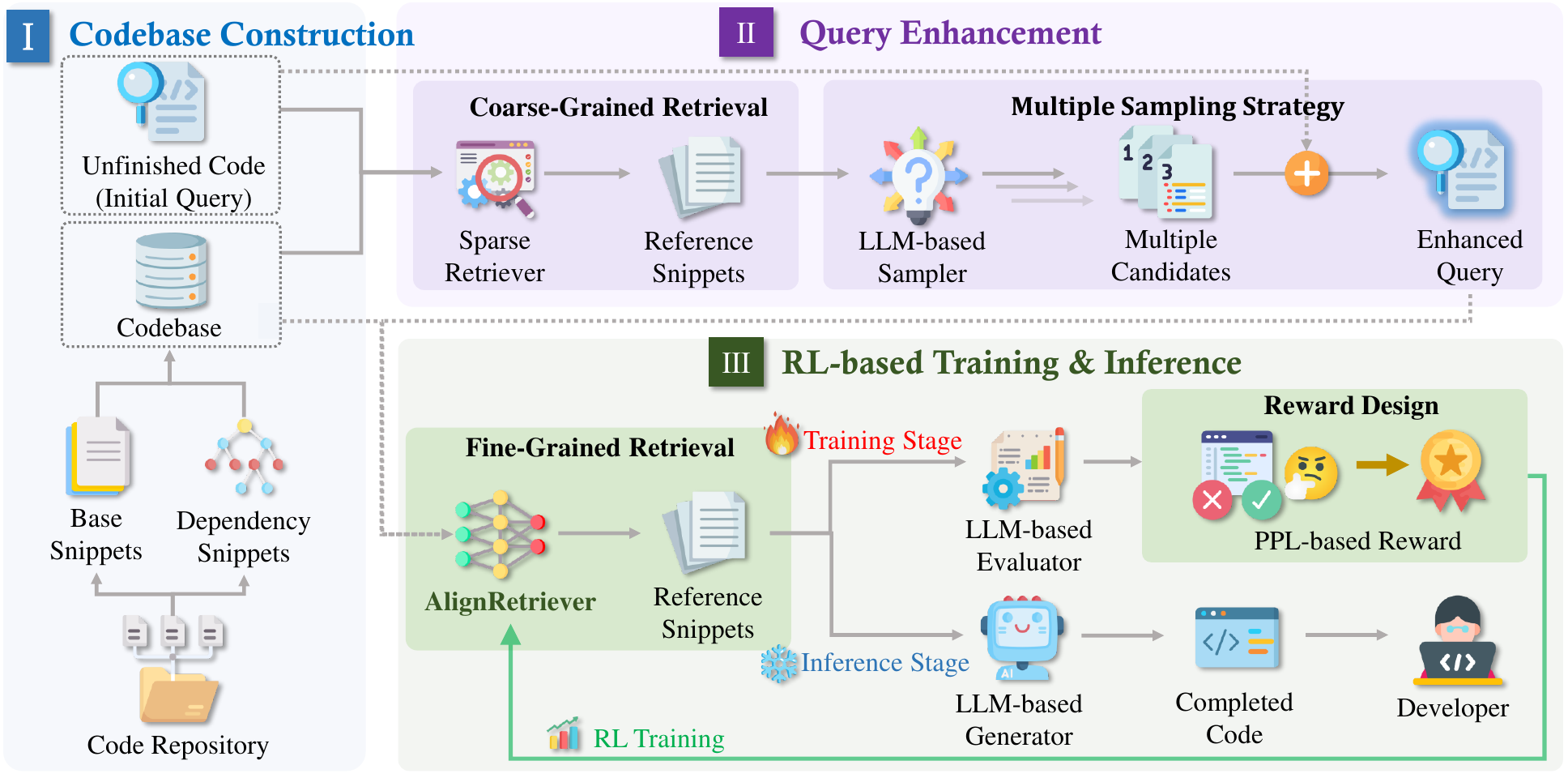}
\vspace{-10pt}
\caption{Overview of AlignCoder. Our approach utilizes the inference capability of the LLM to sample multiple candidate completions and learn to understand the inference information for accurate retrieval for repository-level code completion.} 
\label{fig:overview}
\end{figure*}

In this section, we introduce AlignCoder, a framework for repository-level code completion. 
The pipline of AlignCoder is shown in Figure~\ref{fig:overview}, which contains three stages. \blackcircled{1} Given a code repository, we extract two types of code snippets to construct the codebase for retrieval. \blackcircled{2} For the unfinished code needs to be completed, we use it as the original query to perform the coarse-grained retrieval. Then we use the retrieved code snippets concatenated with the unfinished code as the prompt for sampler. The sampler is a lightweight LLM to sample multiple candidate completions. Then we concatenate the unfinished code with the sampled candidate completions to serve as the enhanced query. \blackcircled{3} We use the enhanced query to perform fine-grained retrieval, which returns several reference code snippets. In the training phase, the reward model will give each reference code snippet a reward to evaluate the helpfulness. The parameter of the retriever will be updated by the reward, which is calculated through an LLM-based evaluator. In the inference phase, we concatenate the retrieved code snippets with the unfinished code to construct the prompt, based on which the generator performs the final code completion.

\begin{figure}[t]
\centering
\includegraphics[width=0.99\columnwidth]{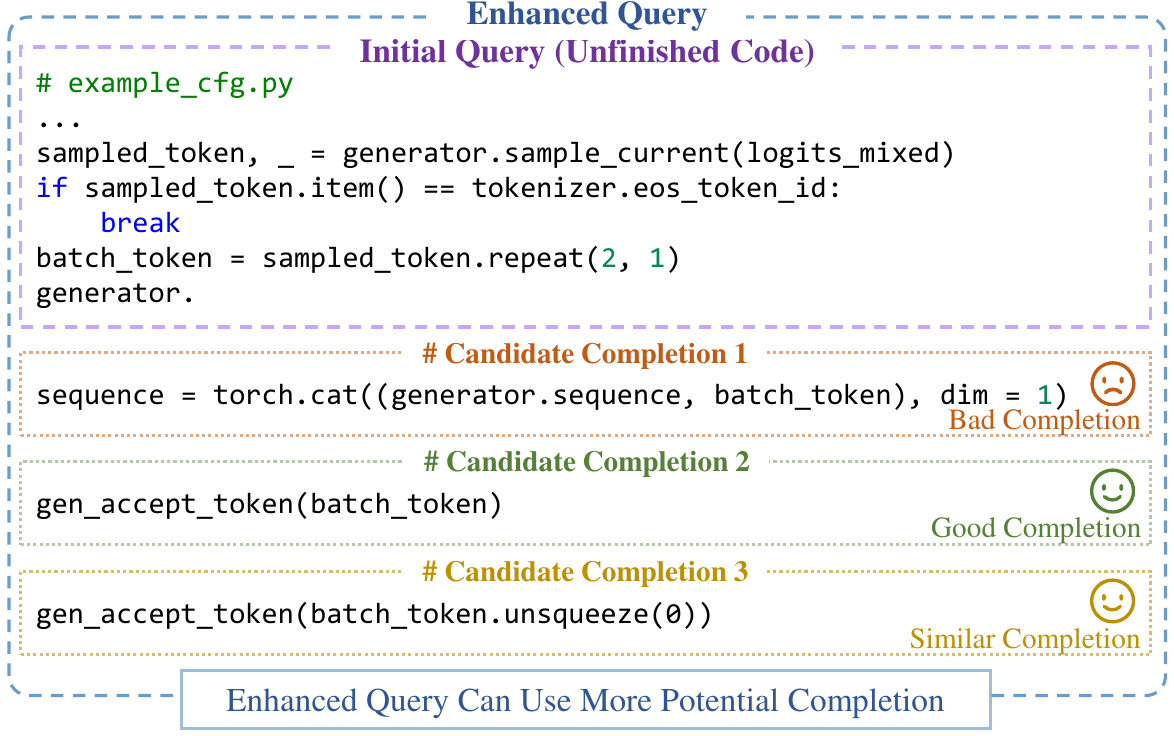}
\vspace{-20pt}
\caption{Example: the initial query and the enhanced query obtained with the query enhancement mechanism. Task id: \texttt{project\_cc\_python/74}.}  
\label{fig:query_enhance_example}
\end{figure}

\subsection{Codebase Construction}
In constructing the retrieval codebase, we construct two distinct types of code snippets, namely base and dependency code snippets. Each category of code snippets is tailored to represent a corresponding type of cross-file information. The first type is the base context, which consists of initial code segments within the cross-files. The second type is dependency context, which provides deep semantic understanding of class hierarchies and API interactions within the codebase~\cite{liang2024repofuse}. The following section provides a detailed introduction to the construction methodologies for base and dependency code snippets.

\subsubsection{Base Code Snippets Construction}

Previous works construct retrieval codebase using fixed window sizes~\cite{zhang2023repocoder} and dependency parsing approaches~\cite{ding2023cocomic, liu2024graphcoder, liu2024stall+}. However, fixed window strategies may disrupt code continuity, while dependency parsing-based methods may only focus on limited context in the context graph and struggle to be applied to complex scenarios, such as when repository dependencies are highly intricate. Therefore, we adopt the Split-Aggregate strategy~\cite{wang2024rlcoder} to construct base code snippets. This approach is inspired by human programming habits, first dividing the cross-files into mini-blocks based on blank lines, then aggregating these mini-blocks into code snippets according to a predefined length. Formally, given a cross-file $F = \{l_1, l_2, \ldots, l_n\}$ where $l_i$ represents the $i$-th line, we first split $F$ into mini-blocks based on blank lines:
\begin{equation}
\text{Split}(F) = \{B_1, B_2, \ldots, B_k\} \text{ where } B_j = \{l_s, \ldots, l_e\}
\end{equation}
\noindent where each mini-block $B_j$ is a contiguous sequence of non-empty lines. Subsequently, we aggregate these mini-blocks into code snippets:
\begin{equation}
S_i = \text{Aggregate}(\{B_j, B_{j+1}, \ldots, B_{j+m}\}) \text{ s.t. } |S_i| \leq L
\end{equation}
\noindent where $S_i$ represents the $i$-th standard code snippet, $L$  is the predefined maximum line counts, and $|S_i|$ denotes the line counts of snippet $S_i$. The aggregation process ensures that:
\begin{equation}
\sum_{k=j}^{j+m} |B_k| \leq L \text{ and } \sum_{k=j}^{j+m+1} |B_k| > L
\end{equation}

\subsubsection{Dependency Code Snippets Construction}

We use tree-sitter~\cite{tree-sitter} to extract import statements from in-file context and parse each statement to obtain module name, entity name, and alias. We filter out standard library and third-party imports, retaining only intra-repository module references. Formally, given import statements $I = \{i_1, i_2, \ldots, i_n\}$:
\begin{equation}
I_{intra} = \{i \in I \mid m(i) \notin (\text{StdLib} \cup \text{ThirdParty})\}
\end{equation}
where $m(i)$ extracts the module name of import $i$. The intra-repository references are categorized into class imports $I_c$, method imports $I_m$, and function imports $I_f$. 

We then parse cross-files using tree-sitter and extract matching code bodies based on these references. We refer to the corresponding code bodies as entities~\cite{liu2024graphcoder, cheng2024dataflow}, denoted by $u$. For function and method imports ($i \in I_f \cup I_m$):
\begin{equation}
\text{Dependency}(i) = \text{Signature}(u_i)
\end{equation}
For class imports ($i \in I_c$):
\begin{equation}
\text{Dependency}(i) = \{\sigma_c, \Sigma_m, \Sigma_{nc}, \Sigma_{nm}\}
\end{equation}
where $\sigma_c$ is the main class signature, $\Sigma_m$ contains all class method signatures, $\Sigma_{nc}$ contains nested class signatures, and $\Sigma_{nm}$ contains nested class method signatures. The complete dependency information is:
\begin{equation}
\mathcal{D} = \bigcup_{i \in I_{intra}} \text{Dependency}(i)
\end{equation}
We construct dependency code snippets based on extracted information, treating each imported class's dependency information as an individual dependency code snippet. For imported methods or functions, we aggregate their signatures into a single snippet. Finally, we combine base code snippets and dependency code snippets to form the retrieval codebase.

\subsection{Query Enhancemant Mechanism}
If the query used for retrieval only contains the unfinished code, it is likely to retrieve code snippets that are similar to them. This raises a gap between the query and the target code, potentially leading to reduced retrieval accuracy. To address this issue, we propose a query enhancement mechanism. 
The details of the mechanism are illustrated below.


Our query enhancement mechanism consists of two phases: coarsed-grained retrieval and multiple sampling strategy. (1) In the coarse-grained retrieval, we use BM25 method to retrieve code snippets from the codebase using the unfinished code as the initial query. 
These code snippets may include both base and dependency code snippets, providing the original code information and deep semantic understanding of class hierarchies and API interactions.
(2) These code snippets are then concatenated with the unfinished code to build a prompt. The sampler samples k candidate completions based on this prompt. Finally, these candidate completions are appended to the initial query to construct an enhanced query, which guides the retriever in the fine-grained retrieval phase. For example, Figure~\ref{fig:query_enhance_example} presents a task \texttt{project\_cc\_python/74} with the initial query and the enhanced query after processing through the query enhancemant mechanism. In the task, we need to complete a specific line in example\_cfg.py where a method is called on the \texttt{generator} object. The \texttt{generator} object is an instance of the \texttt{ExLlamaGenerator} class defined in \texttt{generator.py}, and the target code is a method \texttt{get\_accept\_token} defined in this class. Among the candidate completions sampled by the sampler, candidate completion \textbf{1} is an incorrect completion, while candidate completion \textbf{2} is the correct completion, and candidate completion \textbf{3} is a completion similar to the target code. We consider that candidate completions \textbf{2} and \textbf{3} contain key tokens related to the target code. The enhanced query may improve the ability of the retriever to retrieve more relevant code snippets in the following fine-grained retrieval phase.

\subsection{\alignretriever: RL-based Retriever Aglignment Training}

\subsubsection{Training data construction}
We randomly selected 10,000 Python and Java repositories from GitHub. These repositories contain cross-file dependencies and were created before March 2023 and are not included in our evaluation benchmarks, CrossCodeEval~\cite{ding2023crosscodeeval} and RepoEval~\cite{zhang2023repocoder}, ensuring no data leakage and maintaining the fairness of the evaluation. Constructing training data involves the following three steps: \tikz[baseline=(char.base)]{\node[shape=circle,draw,inner sep=1pt] (char) {1};} We divide each code repository into a series of clusters, where the code files within these clusters have interdependent relationships. We exclude clusters that contain only a single file. \tikz[baseline=(char.base)]{\node[shape=circle,draw,inner sep=1pt] (char) {2};} Then, we perform topological sorting on the remaining clusters. This sorting is based on the in-degree and out-degree relationships between the files. The final sorting result ensures that : The first code file is dependent upon other files in the same cluster, but does not contain dependencies on other files. All subsequent code files depend on one or more other files within the cluster. \tikz[baseline=(char.base)]{\node[shape=circle,draw,inner sep=1pt] (char) {3};}  For each cluster, we randomly select one non-first file to construct the target code for completion. To ensure the preceding context of the target code contains sufficient information, we avoid selecting starting positions of the target code from either the beginning or the end of the current file. The length of the target code is also randomly determined (set to be from 16 to 96 tokens in our experiment), and the entire code segment must lie within the boundaries of the current file.

\subsubsection{The training objective function}
The reward mechanism in reinforcement learning provides the model with environmental feedback, allowing it to progressively learn particular capabilities based on this feedback signal. We define the reward function as $\text{Reward}(\cdot)$, where the reward function needs to be maximized during AlignRetriever training period. The detailed mathematical expression for $\text{Reward}(\cdot)$ is:
\begin{equation}
\text{Reward} = \sum_{i=1}^{n} \mathbb{I}(c_i) \times \log \frac{\exp(s_{i,q})}{\sum_{j=1}^{n} \exp(s_{j,q})}
\end{equation}

\noindent where $q$ is the enhanced query, $C = \{c_1, c_2, \ldots, c_n\}$ is the set of retrieved code snippets, $n$ is the number of code snippets, $\mathbb{I}(c_i)$ is the indicator function denoting correctness of snippet $c_i$, and $s_{i,q} = \cos(\text{emb}(c_i), \text{emb}(q))$ represents the cosine similarity between embeddings of code snippet $c_i$ and query $q$.
The definition of $\mathbb{I}(c_i)$ is:
\begin{equation}
\mathbb{I}(c_i) = \begin{cases}
1, & \text{if } c_i = c_{\text{mp}} \\
0, & \text{otherwise}
\end{cases}
\end{equation}

\noindent where \( c_{\text{mp}} \in C \) and satisfies:
\begin{equation}
\text{PPL}(t|q,c_{\text{mp}}) \leq \text{PPL}(t|q,c_i) \quad \forall i \in \{1,\dots,n\} 
\end{equation}

\noindent where the formula \(\text{PPL}(t|q, c_i)\) is defined as:
\begin{equation}
\text{PPL}(t|x, c_i) = e^{-\frac{1}{L} \sum_{j=1}^{L} \log P(t_j|c_i, q, t_{<j})}
\end{equation}

\noindent where \( L \) denotes the total number of tokens in the target code.

\section{Experimental setup}
In this section, we introduce the benchmarks, backbone models, baseline methods, and evaluation metrics used in our experiments.

\subsection{Benchmarks and Backbone Models} 
In our evaluation, we use two benchmarks for repository-level code completion tasks: CrossCodeEval~\cite{ding2023crosscodeeval} and RepoEval~\cite{zhang2023repocoder}. These two benchmarks have been widely used in previous work~\cite{cheng2024dataflow, wang2024rlcoder, guo2024deepseekcoder, hui2024qwen2, wei2023magicoder}.
\begin{itemize}
  \item \textbf{CrossCodeEval}: CrossCodeEval is a diverse and multilingual code completion benchmark that requires a deep understanding of context across different files in the repository to ensure accurate code completion.
  We conduct the evaluation on Python and Java sets.
  \item \textbf{RepoEval}: RepoEval is allowed for assessments at three granularities, line, API invocation, and function body. In our evaluation, we focus on the line-level and API invocation tasks.
\end{itemize}

In our experiments, we select five code LLMs as generators. These code LLMs have been proven to perform well on repository-level code completion tasks in previous work~\cite{zhang2023repocoder,wang2024rlcoder, liang2024repofuse, phan2024repohyper, wu2024repoformer}. These five code LLMs are: CodeLlama-7B~\cite{rozière2024code}, StartCoder-7B~\cite{li2023starcoder}, StarCoder2-7B~\cite{lozhkov2024starcoder}, DeepSeekCoder-1B~\cite{guo2024deepseekcoder}, and DeepSeekCoder-7B~\cite{guo2024deepseekcoder}.

\subsection{Baseline Methods}
In our experiments, we compare AlignCoder with previous RAG-based methods, encompassing ReACC, RepoCoder, and RLCoder.

\begin{itemize}
  \item \textbf{ReACC}: ReACC~\cite{lu2022reacc} adopt the hybrid retriever~\cite{karpukhin-etal-2020-dense, ma2021replicationstudydensepassage} framework by combining sparse and dense retriever.
  
  \item \textbf{RepoCoder}: 
  RepoCoder~\cite{zhang2023repocoder} is an iterative  retrieval and generation framework, where it searches for the relevant code snippets using the output generated by code LLMs from the previous iteration.
  \item \textbf{RLCoder}:  RLCoder~\cite{wang2024rlcoder} is a  reinforcement learning framework, which can enable the retriever to learn to retrieve useful snippets without the need for labeled data.
\end{itemize}

\subsection{Evaluation Metrics}
Following the established practice~\cite{liu2024graphcoder,cheng2024dataflow, wang2024rlcoder, zhang2023repocoder}, we use two metrics, {Exact Match (EM)} and {Edit Similarity (ES)}, to evaluate code completion  accuracy. 

\begin{table*}[t]
    \centering
    \setlength{\tabcolsep}{7pt}
    \caption{Performance comparison. Superscripted percentages represent the improvement over the corresponding best baseline.}
    \label{table:performance}
    \vspace{-5pt}
    \begin{tabular}{l ll ll ll ll}
    \toprule
    \multirow{2}{*}{\textbf{Method/Model}} & \multicolumn{2}{c}{\textbf{CrossCodeEval (Python)}} & \multicolumn{2}{c}{\textbf{CrossCodeEval (Java)}} & \multicolumn{2}{c}{\textbf{RepoEval (Line)}} & \multicolumn{2}{c}{\textbf{RepoEval (API)}} \\
    \cmidrule(lr){2-3} \cmidrule(lr){4-5} \cmidrule(lr){6-7} \cmidrule(lr){8-9}
    & \textbf{EM} & \textbf{ES} & \textbf{EM} & \textbf{ES} & \textbf{EM} & \textbf{ES} & \textbf{EM} & \textbf{ES} \\
    \midrule
    \ReACCCLSeven & 21.76 & 69.09 & 23.42 & 66.13 & 42.31 & 64.35 & 34.38 & 61.45 \\
    \RepocoderCLSeven & 23.34 & 70.84 & 24.17 & 66.56 & 43.94 & 65.81 & 37.00 & 63.51 \\
    \RlcoderCLSeven  & 26.64  & 72.26 & 26.27 & 67.60  & 46.63  & 67.86  & 37.94  & 64.32 \\
    \cellcolor{blue!5}\textbf{\AligncoderCLSeven} & 
    \cellcolor{blue!5}\textbf{30.13}\mytextsuperscript{ $ \uparrow$13.1\%}    & 
    \cellcolor{blue!5}\textbf{74.29}\mytextsuperscript{ $ \uparrow$2.8\%} & 
    \cellcolor{blue!5}\textbf{28.80}\mytextsuperscript{ $ \uparrow$9.6\%} & 
    \cellcolor{blue!5}\textbf{68.38}\mytextsuperscript{ $ \uparrow$1.2\%} & 
    \cellcolor{blue!5}\textbf{46.96}\mytextsuperscript{ $ \uparrow$0.7\%} & 
    \cellcolor{blue!5}\textbf{67.92}\mytextsuperscript{ $ \uparrow$0.1\%} & 
    \cellcolor{blue!5}\textbf{39.56}\mytextsuperscript{ $ \uparrow$4.3\%} & 
    \cellcolor{blue!5}\textbf{66.01}\mytextsuperscript{ $ \uparrow$2.6\%} \\
    \midrule
    \ReACCSCSeven & 22.33 & 69.60 & 22.16 & 67.80 & 43.81 & 64.83 & 31.94 & 56.00 \\
    \RepocoderSCSeven & 23.15 & 70.71 & 22.53 & 68.22 & 45.69 & 66.90 & 33.44 & 57.81 \\
    \RlcoderSCSeven  & 26.00  & 72.16  & 25.76  & 68.80  & 47.81  & 68.50 & 35.06 & 58.08  \\
    \cellcolor{blue!5}\textbf{\AligncoderSCSeven} & 
    \cellcolor{blue!5}\textbf{30.43}\mytextsuperscript{ $ \uparrow$17.0\%} &
    \cellcolor{blue!5}\textbf{74.33}\mytextsuperscript{ $ \uparrow$3.0\%} & 
    \cellcolor{blue!5}\textbf{28.00}\mytextsuperscript{ $ \uparrow$8.7\%} & 
    \cellcolor{blue!5}\textbf{70.04}\mytextsuperscript{ $ \uparrow$1.8\%} & 
    \cellcolor{blue!5}\textbf{48.38}\mytextsuperscript{ $ \uparrow$1.2\%} & 
    \cellcolor{blue!5}\textbf{68.56}\mytextsuperscript{ $ \uparrow$0.1\%} & 
    \cellcolor{blue!5}\textbf{36.25}\mytextsuperscript{ $ \uparrow$3.4\%} & 
    \cellcolor{blue!5}\textbf{59.49}\mytextsuperscript{ $ \uparrow$2.4\%} \\
    \midrule
    \ReACCSCTSeven & 22.89 & 70.66 & 23.42 & 69.13 & 44.44 & 65.95 & 34.50 & 58.78 \\
    \RepocoderSCTSeven & 24.35 & 71.71 & 23.75 & 69.59 & 45.81 & 67.37 & 36.44 & 59.92 \\
    \RlcoderSCTSeven & 27.47 & 73.39 & 26.69 & 70.35  & 48.63 & 68.59   & 37.75  & 61.08  \\
    \cellcolor{blue!5}\textbf{\AligncoderSCTSeven} & 
    \cellcolor{blue!5}\textbf{31.74}\mytextsuperscript{ $ \uparrow$15.5\%} & 
    \cellcolor{blue!5}\textbf{75.70}\mytextsuperscript{ $ \uparrow$3.1\%} & 
    \cellcolor{blue!5}\textbf{30.43}\mytextsuperscript{ $ \uparrow$14.0\%} & 
    \cellcolor{blue!5}\textbf{72.64}\mytextsuperscript{ $ \uparrow$3.3\%} & 
    \cellcolor{blue!5}\textbf{48.81}\mytextsuperscript{ $ \uparrow$0.4\%} & 
    \cellcolor{blue!5}\textbf{68.82}\mytextsuperscript{ $ \uparrow$0.3\%} & 
    \cellcolor{blue!5}\textbf{38.69}\mytextsuperscript{ $ \uparrow$2.5\%} & 
    \cellcolor{blue!5}\textbf{61.59}\mytextsuperscript{ $ \uparrow$0.8\%} \\
    \midrule
    \ReACCDSCOne & 19.74 & 67.68 & 18.89 & 62.47 & 39.31 & 62.04 & 33.00 & 60.41 \\
    \RepocoderDSCOne & 20.23 & 68.78 & 19.59 & 62.35 & 40.88 & 63.56 & 35.13 & 61.92 \\
    \RlcoderDSCOne & 24.02 & 70.45 & 20.66 & 63.17 & 44.06 & 66.05 & 36.00 & 62.50 \\
    \cellcolor{blue!5}\textbf{\AligncoderDSCOne} & 
    \cellcolor{blue!5}\textbf{28.37}\mytextsuperscript{ $ \uparrow$18.1\%} & 
    \cellcolor{blue!5}\textbf{73.02}\mytextsuperscript{ $ \uparrow$3.6\%} & 
    \cellcolor{blue!5}\textbf{23.24}\mytextsuperscript{ $ \uparrow$12.5\%} & 
    \cellcolor{blue!5}\textbf{64.58}\mytextsuperscript{ $ \uparrow$2.2\%} & 
    \cellcolor{blue!5}\textbf{44.69}\mytextsuperscript{ $ \uparrow$1.4\%} & 
    \cellcolor{blue!5}\textbf{66.74}\mytextsuperscript{ $ \uparrow$1.0\%} & 
    \cellcolor{blue!5}\textbf{37.25}\mytextsuperscript{ $ \uparrow$3.5\%} & 
    \cellcolor{blue!5}\textbf{64.43}\mytextsuperscript{ $ \uparrow$3.1\%} \\
    \midrule
    \ReACCDSCSeven & 23.30 & 70.84 & 22.49 & 66.78 & 45.69 & 66.67 & 38.00 & 65.66 \\
    \RepocoderDSCSeven & 26.98 & 72.96 & 24.96 & 66.52 & 46.38 & 67.51 & 39.31 & 66.29 \\
    \RlcoderDSCSeven & 30.09 & 74.43 & 26.37 & 67.28 & 48.81 & 69.48 & 39.75 & 66.01 \\
    \cellcolor{blue!5}\textbf{\AligncoderSeven} & 
    \cellcolor{blue!5}\textbf{33.92}\mytextsuperscript{ $ \uparrow$12.7\%} & 
    \cellcolor{blue!5}\textbf{76.97}\mytextsuperscript{ $ \uparrow$3.4\%} & 
    \cellcolor{blue!5}\textbf{28.28}\mytextsuperscript{ $ \uparrow$7.2\%} & 
    \cellcolor{blue!5}\textbf{68.31}\mytextsuperscript{ $ \uparrow$1.5\%} & 
    \cellcolor{blue!5}\textbf{49.56}\mytextsuperscript{ $ \uparrow$1.5\%} & 
    \cellcolor{blue!5}\textbf{69.93}\mytextsuperscript{ $ \uparrow$0.6\%} & 
    \cellcolor{blue!5}\textbf{41.88}\mytextsuperscript{ $ \uparrow$5.4\%} & 
    \cellcolor{blue!5}\textbf{67.75}\mytextsuperscript{ $ \uparrow$2.6\%} \\
    \bottomrule
    \end{tabular}
\end{table*}

\subsection{Experimental Details}
We performed all experiments on a machine configured with 2 NVIDIA Tesla A100 GPUs, each with 80 GB of memory. 
\begin{itemize}
  \item \textbf{Training Stage}: In the training stage, we initialize the retriever using UniXcoder and use DeepSeekCoder-1B as the evaluator. We trained the retriever for a total of 20 epochs. Each epoch utilized 3,000 samples for training, and the learning rate was set to 5e-5.
  \item \textbf{Inference Stage}: In the inference stage, since our approach employs multiple sampling, we utilize the vLLM~\cite{kwon2023efficient} framework to accelerate model inference. 
\end{itemize}

\section{Experimental Results}
We aims to answer the following research questions (RQs):
\begin{itemize}
  \item \textbf{RQ1}: How effective is AlignCoder in repository-level code completion?
  \item \textbf{RQ2}: The effectiveness of multiple sampling strategies, and what the optimal sampling numbers should be?
  \item \textbf{RQ3}: What is the contribution of each AlignCoder component to its performance?
  \item \textbf{RQ4}: What is the robustness of AlignCoder's parameter settings?
\end{itemize}

\subsection{RQ1: Overall Performance of AlignCoder}
In this section, we study the effectiveness of AlignCoder compared with three baseline methods. Table~\ref{table:performance} presents the performance comparison of AlignCoder against three baseline methods across multiple benchmarks and backbone models. The results demonstrate that AlignCoder outperforms previous methods across all evaluation settings.

From the perspective of benchmarks, AlignCoder achieves greater improvements on CrossCodeEval compared to RepoEval. For the CrossCodeEval Python, AlignCoder demonstrates substantial performance gains, with EM score improvements ranging from 12.7\% to 18.1\% compared to RLCoder. For the CrossCodeEval Java, AlignCoder maintains its superiority with EM score improvements of 7.2\% to 14.0\% compared to RLCoder across backbone models.

From the perspective of models, results show that maximal improvement is achieved by DeepSeekCoder-1B on CrossCodeEval Python (EM: +18.1\%, ES: +3.6\%). For CrossCodeEval Java, StarCoder2-7B demonstrated the highest improvement (EM: +14.0\%, ES: +3.3\%). Regarding RepoEval performance, DeepSeekCoder-7B showed the most significant gains. It achieved EM score improvements of 1.5\% on line-level tasks and 5.4\% on API-level tasks.

\begin{center}
    \begin{myboxc} \textbf{RQ1 Summary: }
AlignCoder consistently outperforms all baseline methods across different benchmarks and backbone models. The best-performing setting improves 18.1\% on the EM score, with particularly gains on CrossCodeEval Python.
    \end{myboxc}
\end{center}

\subsection{RQ2: Mutiple Sampling Strategy}
In this section, we investigate two questions: (1) whether multiple sampling is more effective than single sampling, and (2) what the optimal sampling numbers for AlignCoder. We conduct performance comparison experiments with sampling numbers set from 1 to 6.
The generator used in this experiment is DeepSeekCoder-1B. The experimental results are presented in Table~\ref{sampling number}.

\begin{table*}[t]
    \centering
    \setlength{\tabcolsep}{2.2pt}
    \caption{Superscripted percentages indicate the relative increase or decrease of multi-sampling compared to single-sampling.} 
    \label{sampling number}
    \vspace{-5pt}
    \begin{tabular}{lllllllllll}
    \toprule
    \multirow{2}{*}{\textbf{Method}} & \multicolumn{2}{c}{\textbf{CrossCodeEval (Python)}} & \multicolumn{2}{c}{\textbf{CrossCodeEval (Java)}} & \multicolumn{2}{c}{\textbf{RepoEval (Line)}} & \multicolumn{2}{c}{\textbf{RepoEval (API)}} & \multicolumn{2}{c}{\textbf{Average}} \\
    \cmidrule(lr){2-3} \cmidrule(lr){4-5} \cmidrule(lr){6-7} \cmidrule(lr){8-9} \cmidrule(lr){10-11}
    & \textbf{EM} & \textbf{ES} & \textbf{EM} & \textbf{ES} & \textbf{EM} & \textbf{ES} & \textbf{EM} & \textbf{ES} & \textbf{EM} & \textbf{ES} \\
    \midrule
    \sam{1} & 27.20 & 71.98 & 23.05 & 64.49 & 44.75 & 66.47 & 36.69 & 63.39 & 32.92 & 66.58 \\
    \sam{2} & 27.65\mytextsuperscript{ $\uparrow$1.7\%}  & 72.41\mytextsuperscript{ $\uparrow$0.6\%}  & \textbf{23.42}\mytextsuperscript{ $\uparrow$1.6\%}  & \textbf{65.45}\mytextsuperscript{ $\uparrow$1.5\%}  & \textbf{45.19}\mytextsuperscript{ $\uparrow$1.0\%}  & \textbf{66.98}\mytextsuperscript{ $\uparrow$0.8\%}  & 36.63\mytextsuperscript{ $\downarrow$0.2\%}  & 63.90\mytextsuperscript{ $\uparrow$0.8\%} & 33.22\mytextsuperscript{ $\uparrow$0.9\%} & 67.19\mytextsuperscript{ $\uparrow$0.9\%} \\
    \sam{3} &  28.29\mytextsuperscript{ $\uparrow$4.0\%} & 72.71\mytextsuperscript{ $\uparrow$1.0\%}  & 23.19\mytextsuperscript{ $\uparrow$0.6\%} & 63.93\mytextsuperscript{ $\downarrow$0.9\%}  & 44.75\mytextsuperscript{ $\uparrow$0.0\%}  & 66.35\mytextsuperscript{ $\downarrow$0.2\%}  & 36.88\mytextsuperscript{ $\uparrow$0.5\%}  & 63.93\mytextsuperscript{ $\uparrow$0.9\%} & 33.28\mytextsuperscript{ $\uparrow$1.1\%} & 66.73\mytextsuperscript{ $\uparrow$0.2\%} \\
    \cellcolor{blue!8}\textbf{\sam{4}} & \cellcolor{blue!8}28.37\mytextsuperscript{ $\uparrow$4.3\%} & \cellcolor{blue!8}\textbf{73.02}\mytextsuperscript{ $\uparrow$1.5\%} & \cellcolor{blue!8}23.24\mytextsuperscript{ $\uparrow$0.8\%} & \cellcolor{blue!8}64.58\mytextsuperscript{ $\uparrow$0.1\%} & \cellcolor{blue!8}44.69\mytextsuperscript{ $\uparrow$-0.1\%} & \cellcolor{blue!8}66.74\mytextsuperscript{ $\uparrow$0.4\%} & \cellcolor{blue!8}37.25\mytextsuperscript{ $\uparrow$1.5\%} & \cellcolor{blue!8}\textbf{64.43}\mytextsuperscript{ $\uparrow$1.6\%} & \cellcolor{blue!8}\textbf{33.39}\mytextsuperscript{ $\uparrow$1.4\%} & \cellcolor{blue!8}\textbf{67.19}\mytextsuperscript{ $\uparrow$0.9\%} \\
    \sam{5} &\textbf{28.44}\mytextsuperscript{ $\uparrow$4.6\%}  &71.68\mytextsuperscript{ $\downarrow$0.4\%}  &22.81\mytextsuperscript{ $\downarrow$1.0\%}  &64.10\mytextsuperscript{ $\downarrow$0.6\%}  &44.37\mytextsuperscript{ $\downarrow$0.9\%}  &66.44\mytextsuperscript{ $\downarrow$0.1\%}  &36.63\mytextsuperscript{ $\downarrow$0.2\%}  &63.43\mytextsuperscript{ $\uparrow$0.1\%} & 33.06\mytextsuperscript{ $\uparrow$0.4\%} & 66.41\mytextsuperscript{ $\downarrow$0.3\%} \\
    \sam{6} &28.07\mytextsuperscript{ $\uparrow$3.2\%}  &72.54\mytextsuperscript{ $\uparrow$0.8\%}  &22.87\mytextsuperscript{ $\downarrow$0.8\%}  &64.18\mytextsuperscript{ $\downarrow$0.5\%}  &44.63\mytextsuperscript{ $\downarrow$0.3\%}  &66.38\mytextsuperscript{ $\downarrow$0.1\%}  &\textbf{37.81}\mytextsuperscript{ $\uparrow$3.1\%}  &64.07\mytextsuperscript{ $\uparrow$1.1\%} & 33.35\mytextsuperscript{ $\uparrow$1.3\%} & 66.79\mytextsuperscript{ $\uparrow$0.3\%} \\
    \bottomrule
    \end{tabular}
\end{table*}

\textbf{(1) Effectiveness of Multiple Sampling.} 
Multiple sampling (2, 3, and 4 samples) generally outperforms single sampling in EM and ES scores. 
The few exceptions show only minimal decreases: 0.2\% lower EM on API-level tasks of RepoEval (sampling numbers set to 2), for instance.
These results indicate that multiple sampling enhances the likelihood of generating key tokens related to target code, thus improving the retrieval performance.

\textbf{(2) Performance degradation in certain datasets when sampling numbers exceed a threshold.}
A performance degradation is observed on certain datasets when sampling numbers surpass 4. When the sampling number is set to 5, the results on CrossCodeEval Java and line-level/API-level tasks of RepoEval are all inferior to single sampling. When the sampling number is set to 6, there is a decline compared to single sampling on CrossCodeEval Java and line-level tasks of RepoEval. This indicates that although multiple sampling is effective compared to single sampling, excessive sampling may cause the candidate completions to contain non-negligible noise, which could affect the accuracy of subsequent retrieval. 

\textbf{(3) Determining the optimal sampling numbers.}
We evaluate the performance of AlignCoder by computing average EM and ES scores under different sampling settings. As shown in Table~\ref{sampling number}, the setting of sampling numbers at 4 demonstrates superior performance. This evidence establishes sampling numbers of 4 as the optimal balance for our approach, maximizing performance gains while maintaining computational efficiency.

\begin{center}
    \begin{myboxc} \textbf{RQ2 Summary: }
We demonstrate the effectiveness of the multiple sampling strategy compared to single sampling. 
When the sampling number exceeds a threshold, AlignCoder's performance declines. Experimental results show that the optimal sampling number is 4.
    \end{myboxc}
\end{center}

\vspace{-5pt}

\begin{table}[t]
    \centering
    \vspace{-10pt}
    \setlength{\tabcolsep}{7pt}
    \renewcommand{\arraystretch}{1.1}
    \begin{threeparttable}
    \caption{Ablation study results.}
    \label{Ablation_Study}
    \vspace{-5pt}
    \begin{tabular}{llllll}
    \toprule
    \multirow{2}{*}{\textbf{Method}} & \multicolumn{2}{c}{\textbf{CrossCodeEval (Python)}} & \multicolumn{2}{c}{\textbf{CrossCodeEval (Java)}} \\
    \cmidrule(lr){2-3} \cmidrule(lr){4-5}
    & \textbf{EM} & \textbf{ES} & \textbf{EM} & \textbf{ES} \\
    \midrule
    AlignCoder & 33.92 & 76.97 & 28.28 & 68.31 \\
    \ \ w/o DC & 31.44\textsuperscript{$\downarrow$7.3\%} & 75.82\textsuperscript{$\downarrow$1.5\%} & 27.44\textsuperscript{$\downarrow$3.0\%} & 67.54\textsuperscript{$\downarrow$1.1\%} \\
    \ \ w/o QH & 31.33\textsuperscript{$\downarrow$7.6\%} & 75.21\textsuperscript{$\downarrow$2.3\%} & 27.12\textsuperscript{$\downarrow$4.1\%} & 67.63\textsuperscript{$\downarrow$1.0\%} \\
    \ \ w/o RL & 28.14\textsuperscript{$\downarrow$17.4\%} & 73.39\textsuperscript{$\downarrow$4.7\%} & 25.62\textsuperscript{$\downarrow$9.4\%} & 66.85\textsuperscript{$\downarrow$2.1\%} \\
    \cmidrule(lr){1-6}
    \multirow{2}{*}{\textbf{Method}} & \multicolumn{2}{c}{\textbf{RepoEval (Line)}} & \multicolumn{2}{c}{\textbf{RepoEval (API)}} \\
    \cmidrule(lr){2-3} \cmidrule(lr){4-5}
    & \textbf{EM} & \textbf{ES} & \textbf{EM} & \textbf{ES} \\
    \midrule
    AlignCoder & 49.56 & 69.93 & 41.88 & 67.75 \\
    \ \ w/o DC & 49.06\textsuperscript{$\downarrow$1.0\%} & 69.96\textsuperscript{$\uparrow$0.0\%} & 40.44\textsuperscript{$\downarrow$3.4\%} & 66.75\textsuperscript{$\downarrow$1.5\%} \\
    \ \ w/o QH & 48.94\textsuperscript{$\downarrow$1.3\%} & 69.63\textsuperscript{$\downarrow$0.4\%} & 40.19\textsuperscript{$\downarrow$4.0\%} & 66.52\textsuperscript{$\downarrow$1.8\%} \\
    \ \ w/o RL & 46.56\textsuperscript{$\downarrow$6.1\%} & 68.06\textsuperscript{$\downarrow$2.7\%} & 38.63\textsuperscript{$\downarrow$7.8\%} & 65.10\textsuperscript{$\downarrow$3.9\%} \\
    \bottomrule
    \end{tabular}
    \begin{tablenotes}
    \small
    \item
    \end{tablenotes}
    \end{threeparttable}
\end{table}

\begin{figure}[t]
\centering
\vspace{-10pt}
\includegraphics[width=\linewidth]{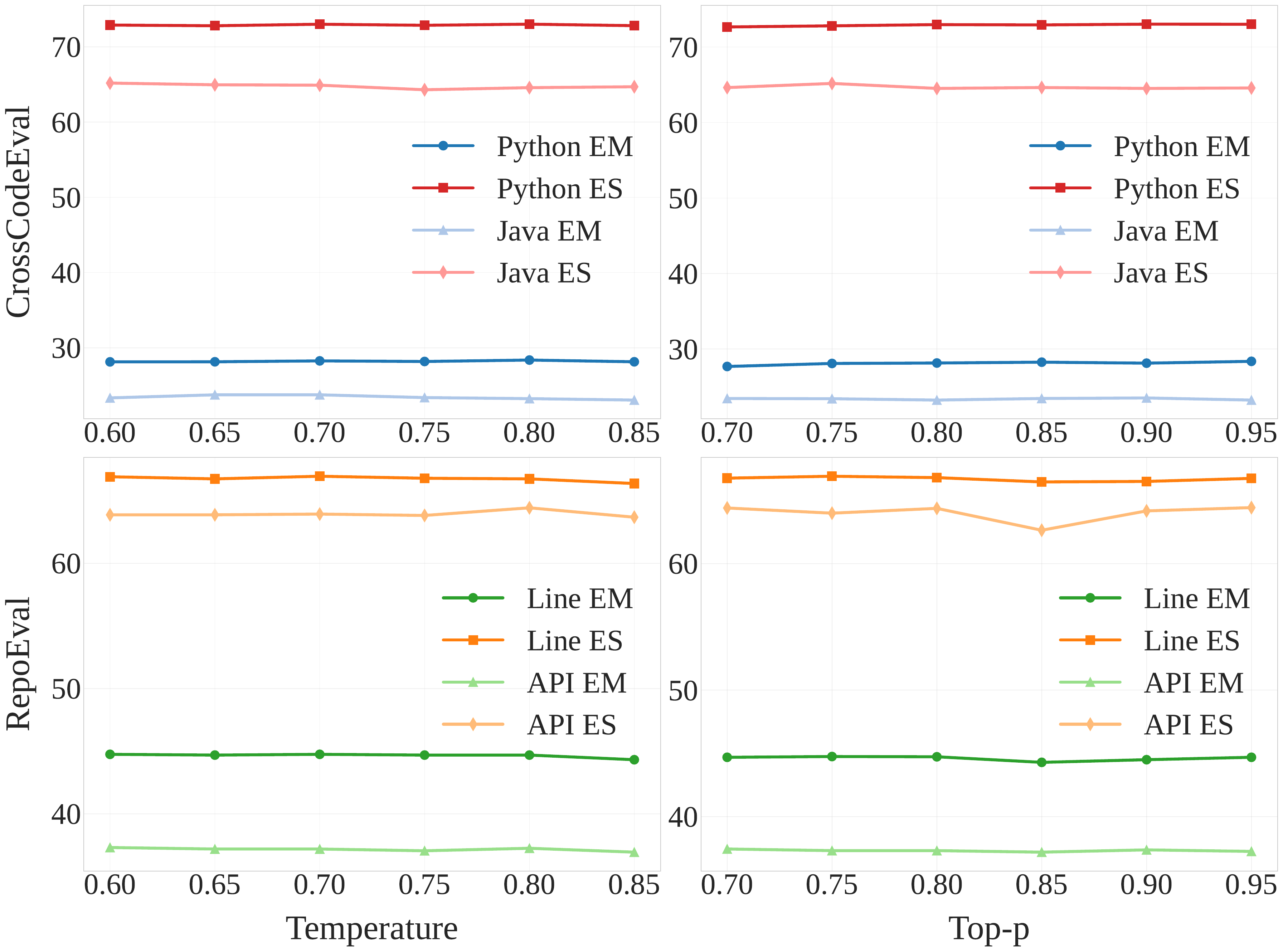}
\vspace{-20pt}
\caption{Temperature and top-$p$ sampling parameter effects on AlignCoder performance.} 
\label{fig:parameter_stability}
\end{figure}

\subsection{RQ3: Ablation Studies}
To demonstrate the effectiveness of the three core components in AlignCoder, we conduct experiments focusing on: (1) incorporating dependency context as cross-file information by constructing and retrieving dependency code snippets; (2) employing a query enhancement mechanism; and (3) using reinforcement learning to train the retriever to learn to utilize the inference information in the enhanced query. We conduct ablation studies on the CrossCodeEval and RepoEval. All experiments are implemented with DeepSeekCoder-7B as the generator, with detailed results shown in Table~\ref{Ablation_Study}.

Table~\ref{Ablation_Study} shows three ablation settings. w/o DC indicates that the dependency context is not considered. w/o QH denotes the configuration without the query enhancement mechanism applied. w/o RL indicates that reinforcement learning is not applied to the retriever, so the retriever has not learned how to utilize the key tokens in the enhanced query.

Our ablation study reveals performance patterns: (1) when w/o DC and w/o QH, both EM and ES scores consistently decrease across benchmarks. The only exception is a negligible 0.08\% improvement in ES for RepoEval line-level tasks in the w/o DC setting. (2) Performance degradation becomes more pronounced under w/o RL. The experimental results show that w/o RL has a significant impact on AlignCoder's performance, which demonstrates that merely considering dependency context and query enhancement mechanism is insufficient, and it is also necessary to train the retriever to learn to utilize the additional information in the query.

\begin{center}
    \begin{myboxc} \textbf{RQ3 Summary: }
Our ablation studies demonstrate the critical roles of three components. Results show that w/o RL causes greater performance drops. This indicates that dependency context and query enhancement are incomplete without training the retriever to exploit the additional information in the enhanced query.
    \end{myboxc}
\end{center}

\begin{figure*}[t]
\centering
\includegraphics[width=\linewidth]{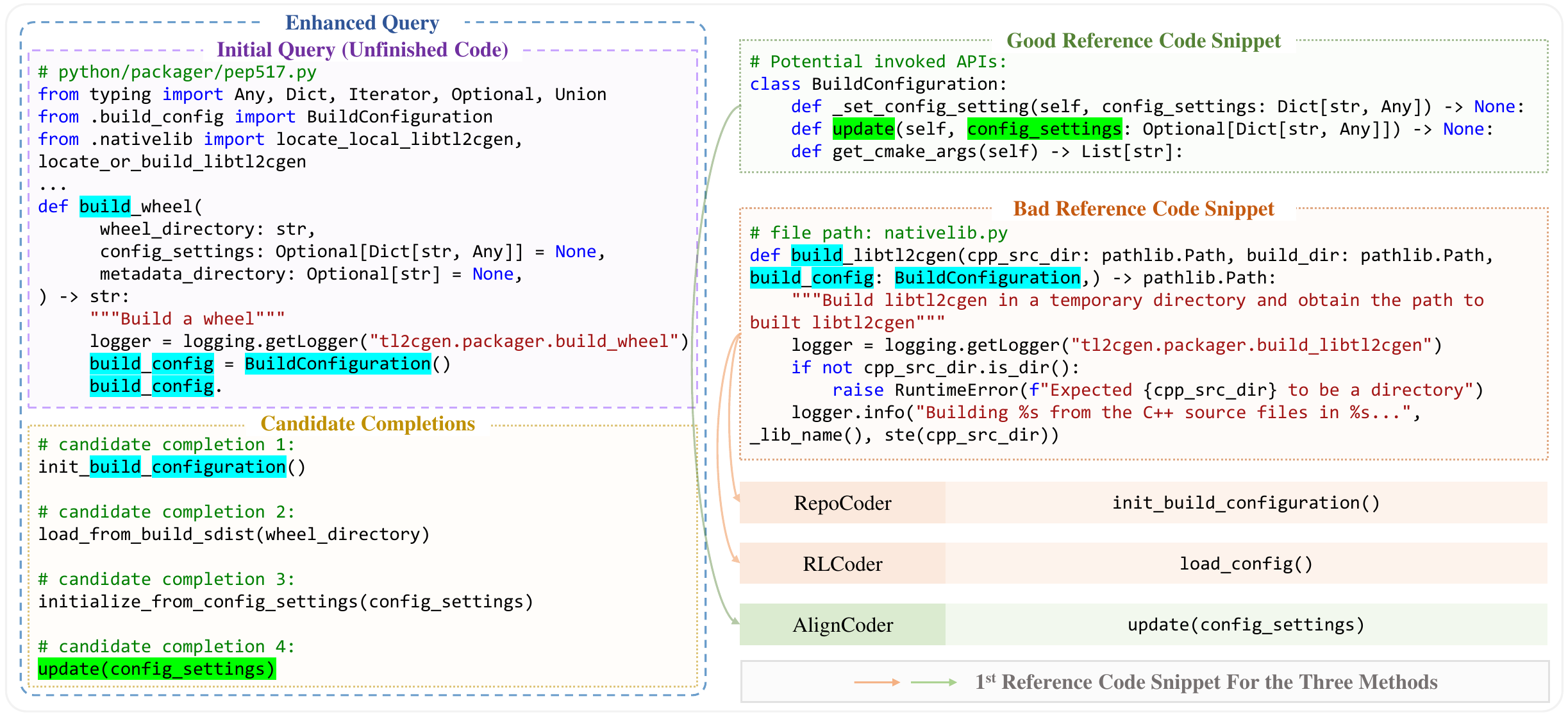}
\vspace{-20pt}
\caption{Case study, task id:  \textit{project\_cc\_python/5407}. The highlighted words represent identical tokens that exist in both the query and the reference code.} 
\label{fig:case_study_1}
\end{figure*}

\subsection{RQ4: Sampling Parameter Stability}

In this section, we analyze the influence of two sampling parameters, temperature and top-$p$, on AlignCoder's performance when sampling candidate completions. Temperature and top-$p$ are two crucial parameters that control the randomness and diversity of generation. Temperature regulates the degree of randomness in generated text, while top-$p$ constrains the candidate vocabulary by sampling only from the highest-probability tokens whose cumulative probability reaches the specified $p$ value.
Since we employ the vLLM framework to accelerate model inference, the temperature and top-$p$ settings in AlignCoder follow the default parameter configurations for the model sampling process as specified in the vLLM documentation, with temperature set to 0.8 and top-$p$ set to 0.95. We conducted a series of experiments to examine the stability of AlignCoder's performance varying temperature (0.85, 0.75, 0.7, 0.65, 0.6) and top-$p$ (0.9, 0.85, 0.8, 0.75, 0.7) values when using the optimal sampling number 4.

Our experiments used DeepSeekCoder-1B as the generator, Figure~\ref{fig:parameter_stability} demonstrates that parameter adjustments typically lead to modest performance changes. Most configurations exhibit EM and ES score variations of less than 1\% across both benchmarks. However, five configurations show differences: a 2.4\% EM decrease on CrossCodeEval Python under temperature at 0.8 and top-$p$ at 0.7, for instance.

\begin{center}
    \begin{myboxc} \textbf{RQ4 Summary: }
AlignCoder demonstrates robust performance under different temperature and top-p sampling configurations, consistently exhibiting stability in EM and ES metrics across most experimental configurations.
    \end{myboxc}
\end{center}

\subsection{Case Study}
 We illustrate the effectiveness of AlignCoder through a case study. In the Figure~\ref{fig:case_study_1}, the model needs to complete the function \texttt{update} of the \texttt{BuildConfiguration} class. The highest similarity code snippet retrieved through AlignCoder is a dependency code snippet related to the \texttt{Validator} class. Based on this snippet, the generator is able to generate the correct completion. However, the most similar code snippet retrieved by RLCoder or RepoCoder is a base code snippet highly similar to the unfinished code, ultimately leading the model to produce an incorrect result.

\section{Related Work}

\subsection{Code Completion}
\revised{In the field of intelligent software engineering, large language models (LLMs) and agents are leveraged to address complex development challenges~\cite{zheng2025towards,yang2025large,zheng2023survey,wang2025towards,wang2025agents, zhou2025adaptive, chen2024identifying, yang2024hyperion}. The field has witnessed significant progress in code generation~\cite{shi2023sotana,li2024repomincoder,wang2025beyond,zheng2024top,zhang2025llm,li2025s,quan2025codeelo,si2025design2code,gu2025retrieve,chen2024rmcbench,zheng2024humanevo,wang2021code,wang2024rlcoder,lai2025analogcoder, zhu2025domaineval,nie2023unveiling}, code search~\cite{gu2024secret,gu2025spencer,gong2025cosqa+,gu2022accelerating,hu2023revisiting,shi2023cocosoda,li2023rethinking,chen2023needs,wang2023you,hu2024tackling,dong2024improving,zheng2024costv, li2025search, chi2024empirical, wang2022enriching,chen2024decoder,zhang2023code}, issue resolution~\cite{guo2025omnigirl,guo2025swe,tao2024magis, chen2025swe, ma2025alibaba,li2025swe, xie2025swe, chen2025prometheus}, code summarization~\cite{shi2022evaluation, shi2021cast}, code translation~\cite{wang2024repotransbench,ou2024repository, pan2024lost, tao2024unraveling, yan2023codetransocean}, commit message generation~\cite{tao2022large,tao2024kadel,xue2024automated,zhang2024using,zhang2024automatic,tao2021evaluation,shi2022race,guo2023snippet,zhang2023ealink}, efficient model optimization~\cite{wang2024sparsecoder,guo2024stop, guo2023longcoder,gim2024prompt, cai2024pyramidkv, yue2024wkvquant, feng2024ada}, and code understanding tasks~\cite{zhang2023code,bai2024longbench,wang2021cocosum,liao2025e2llm,tao2021evaluation, li2025deepcircuitx, wang2025towards, zhou2025adaptive}. These methodologies have collectively advanced our understanding of how AI systems can effectively support software engineering tasks across various domains.}
Auto code completion has been a fundamental task in intelligent software engineering, aiming to predict subsequent code tokens or statements to assist programmers during development~\cite{izadi2024language,wang2023practitioners}. Traditional approaches relied on rule-based methods or statistical methods~\cite{hindle2016naturalness, raychev2014code, bruch2009learning, robbes2008program}, but recent advances have been driven by deep learning techniques. Modern neural approaches~\cite{bhoopchand2016learning,karampatsis2020big, proksch2015intelligent} have significantly improved completion accuracy by learning from large codebases. The emergence of large language models has further revolutionized this field, with studies exploring LLM-based completion systems~\cite{chen2021evaluating,wang2023codet5plus}. Retrieval-augmented generation (RAG) has become particularly influential, with methods like RedCoder~\cite{parvez2021rag} enhancing code generation relevant code snippets retrieval, DocPrompting~\cite{zhou2023docpromptingrag} leveraging documentation for unseen functions, and AceCoder~\cite{li2023acecoder} integrating examples retrieval with guided generation to improve completion quality.

\subsection{Repository-Level Code Completion}

Repository-level code completion extends traditional completion by leveraging broader repository context to improve accuracy and relevance~\cite{ding2023cocomic,phan2024repohyper,bairi2023codeplan,liang2024repofuse,liao2024a3codgen,zhang2023repocoder}. Existing approaches can be categorized into several paradigms. Learning-based methods such as CoCoMIC~\cite{ding2023cocomic} and RepoHyper~\cite{phan2024repohyper} employ dependency analysis and adaptive learning mechanisms, though they face challenges in training data acquisition and generalizability. Static analysis approaches, including CodePlan~\cite{bairi2023codeplan}, RepoFuse~\cite{liang2024repofuse}, and $A^3\text{-CodeGen}$~\cite{liao2024a3codgen}, utilize program analysis techniques to identify relevant code candidates. GraphCoder~\cite{liu2024graphcoder} captures the context by leveraging the structural information in the source code via a constructed code context graph. Iterative retrieval strategies have been explored by RepoCoder~\cite{zhang2023repocoder}, which performs multi-round retrieval and generation, and De-Hallucinator~\cite{eghbali2024dehallucinator}, which refines completion through iterative processes. Tool-augmented methods such as CodeAgent~\cite{zhang2024codeagent} and ToolGen~\cite{wang2024teaching} investigate external tool invocation, while RLCoder~\cite{wang2024rlcoder} employs reinforcement learning for repository-specific retrieval optimization. 


Despite these advances, \aligncoder differs by leveraging LLMs' powerful inference capabilities to obtain multiple reference code snippets, combining this with a reinforcement learning-based approach for training the retriever to achieve more accurate retrieval and superior repository-level code completion performance.

\section{conclusion}
In this paper, we present \aligncoder, a repository-level code completion framework that addresses the fundamental challenges of query-target misalignment and ineffective inference information utilization in existing retrieval-augmented generation approaches. Our key contributions include a query enhancement mechanism that generates multiple candidate completions to bridge the semantic gap between queries and target codes, and a reinforcement learning-based Align-retriever that learns to leverage inference information for more precise retrieval. Extensive experiments on CrossCodeEval and RepoEval benchmarks across five backbone models demonstrate the effectiveness of our approach, achieving an 18.1\% improvement in EM score on the Python subset of CrossCodeEval and showing strong generalizability across various code LLMs and programming languages. This work contributes to the development of repository-level code generation tools and may help improve developer productivity in programming tasks.

\section*{Acknowledgements}

This work is supported by CCF-Huawei Populus Grove Fund CCF-HuaweiSE202403.

\bibliographystyle{IEEEtran}
\bibliography{ref}

\end{document}